# Practical Timing Closure in FPGA and ASIC Designs: Methods, Challenges, and Case Studies

Mostafa Darvishi, *Member, IEEE*

*Abstract*—This paper presents an in-depth analysis of timing closure challenges and constraints in Field Programmable Gate Arrays (FPGAs) and Application Specific Integrated Circuits (ASICs). We examine core timing principles, architectural distinctions, and design methodologies influencing timing behavior in both technologies. A case study comparing the Xilinx Kintex UltraScale+ FPGA (XCKU040) with a 7nm ASIC highlights practical timing analysis and performance trade-offs. Experimental results show ASICs achieve superior timing—45ps setup and 35ps hold—while modern FPGAs remain competitive with 180ps setup and 120ps hold times, validating their suitability for high-performance designs.

## I. INTRODUCTION

TIMING constraints are a fundamental bottleneck in modern digital design, directly affecting maximum operating frequency and system performance. Setup and hold time violations remain critical sources of functional failure in synchronous circuits. As technology scales and clock speeds rise, effective management of these constraints becomes essential for reliable implementation [1]. With continuous advances in FPGA technology narrowing the performance gap with ASICs, a detailed comparative analysis of their timing characteristics is increasingly relevant. This paper explores setup and hold-time behavior in both platforms, highlighting architectural differences and design strategies. [2], [3].

## II. BACKGROUND

Digital timing analysis has advanced considerably since the inception of integrated circuits. The concepts of setup and hold time were formally introduced in the 1970s alongside the rise of synchronous systems. Initially, timing verification relied on exhaustive simulation, but as design complexity increased, this approach became impractical, driving the development of more scalable analytical and static timing analysis techniques. [4]. he introduction of Static Timing Analysis (STA) in the 1980s revolutionized digital design by enabling simulation-free timing verification. STA analyzes all timing paths to detect potential violations, ensuring that setup and hold constraints are met across all possible input scenarios without requiring exhaustive functional simulation. This approach significantly improved scalability and efficiency, making it the standard method for timing validation in modern digital systems. [5]. Systematic Comparison between FPGA and ASIC performance began in the late 1990s, initially emphasizing area and power metrics, with timing as a secondary concern. The seminal work by Rose et al. (1999) laid the groundwork for architectural comparisons between reconfigurable and fixed-function logic, shaping subsequent research in the field. [6]. This paper presents experimental measurements comparing a 90nm CMOS FPGA with 90nm CMOS standard-cell ASICs in terms of logic density, speed, and power consumption. The study by Kuon and Rose (2007) remains one of the most comprehensive analyses, setting methodological standards still used today [7], [8]. By implementing identical circuits on both platforms and measuring actual performance, their work demonstrated that FPGAs incur 3–4× longer delays, 5–15× greater area usage, and 5–14× higher power consumption than ASICs—establishing baseline metrics for ongoing comparative research. [9]. Recent research has explored key areas that enhance timing optimization in FPGAs. One emerging trend is the use of machine learning to predict timing delays during FPGA placement, as demonstrated in [10], which aims to improve timing closure efficiency and reduce design iterations. Environmental considerations have also gained attention, with new static timing analysis techniques incorporating real operating conditions—particularly valuable in harsh environments like aerospace and automotive systems. Additionally, hybrid architecture that combines the efficiency of ASICs with the flexibility of FPGAs are being investigated, such as designs using fine-grained arrays of dynamically reconfigurable processing elements. Before delving into specific technical details, it is essential to emphasize that both Field-Programmable Gate Arrays (FPGAs) and Application-Specific Integrated Circuits (ASICs) share foundational timing principles—such as setup and hold constraints, clock skew, and path delays. However, despite this common ground, the timing analysis methodologies for these technologies differ significantly due to their distinct architectural characteristics. ASICs benefit from a fixed, deterministic hardware fabric, enabling precise and predictable timing closure during the design phase. In contrast, FPGAs introduce additional complexity stemming from their programmable interconnects, configurable logic blocks, and tool-dependent routing strategies. These characteristics introduce unique timing uncertainties and demand specialized analysis and optimization techniques tailored to reconfigurable platforms. The following subsection elaborates on the timing challenges inherent to FPGA-based designs, followed by a discussion on timing complexities in ASIC design [11], [12].

**Timing Challenges in FPGA Design**:

Mostafa Darvishi is with the department of Electrical Engineering at École de technologie supérieure (ÉTS). He is also the VP of Engineering at Evolution Optiks Limited (darvishi@ieee.org).

- *Configuration-Dependent Timing*: Traditional cell timing models using the Liberty format cannot account for timing variations introduced by different cell configurations. This limitation has prompted the development of FPGA- specific timing models that incorporate the effects of reconfigurability.
- *Routing Architecture Impact*: The hierarchical and programmable routing fabric of FPGAs introduces variability in interconnect delays. Unlike ASICs, which have fixed routing, FPGAs require statistical timing models that capture routing uncertainties, especially during early design stages.
- *Place-and-Route Interdependency*: FPGA timing analysis is tightly coupled with placement and routing. To address this, new timing-driven detailed placement techniques such as those optimizing critical paths are wssential for improving accuracy and closure efficiency.

**Timing Challenges in ASIC Design** [13]–[15]:

- *Process Variation Modeling*: At advanced nodes, statistical modeling is critical to capture intra-die and inter-die manufacturing variations. This has led to the widespread adoption of statistical static timing analysis (SSTA).
- *Physical Effects Integration*: Modern ASIC flows incorporate physical phenomena like crosstalk, IR drop, and temperature gradients into the timing analysis process, particularly relevant at 7nm and below.
- *Hierarchical Timing Analysis*: Given the scale of modern ASICs, hierarchical analysis frameworks are necessary to manage millions of timing paths while preserving accuracy and computational efficiency.

Timing constraints have evolved from simple clock period definitions to complex multi-mode, multi-corner (MMMC) scenarios, reflecting modern design complexity. Industry-standard formats like Synopsys Design Constraints (SDC) enable portable, tool-independent constraint specification [16]. Static Timing Analysis (STA) remains the primary method for timing closure verification by analyzing all timing paths without input vectors. However, as designs grow more complex and adaptive, hybrid methods combining STA with dynamic validation have emerged to improve timing confidence [17]. With CMOS scaling to nanometer nodes, timing analysis complexity has increased due to interconnect delays dominating gate delays—especially in FPGAs with extensive programmable routing [18]. Process variations at advanced nodes have reduced the accuracy of worst-case corner analysis, prompting the adoption of Statistical Static Timing Analysis (SSTA) for more realistic, probabilistic timing predictions [19].

## III. SOME THEORETICAL BACKGROUND

### A. Setup Time Fundamentals

Setup time ($T_{setup}$) is the minimum duration that data must be stable before the clock's active edge to ensure correct capture by a flip-flop. This allows data to propagate and settle through input logic prior to sampling. It can be expressed mathematically as [20]–[22]:

$$T_{dataArrival} + T_{setupRequirement} \leq T_{clockArrival} \quad (1)$$

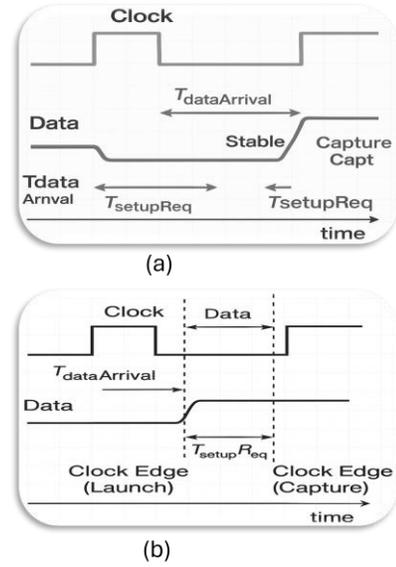

Figure 1. Setup time representation in a digital system

Where:
$T_{dataArrival}$: Time data arrives at the flip-flop input
$T_{setupRequirement}$: Minimum required setup time of the flip-flop
$T_{clockArrival}$: Arrival time of the clock edge at the flip-flop
A setup violation occurs if:

$$T_{dataArrival} + T_{setupRequirement} > T_{clockArrival} \quad (2)$$

which causes incorrect data capture. Figure 1(a) shows the timing diagram representation of setup time.

### B. Hold Time Fundamentals

Hold time, $T_{hold}$, is the minimum time the data input must remain stable after the active clock edge to ensure correct data capture. This constraint can be expressed as [20], [23]–[25]:

$$T_{dataStableDuration} \geq T_{holdRequirement} \quad (3)$$

Where:
$T_{dataStableDuration}$: Duration data remains stable after the clock edge
$T_{holdRequirement}$: Minimum hold time specified for the flip-flop
A hold violation occurs if:

$$T_{dataStableDuration} < T_{holdRequirement} \quad (4)$$

which can cause metastability or data corruption.

Figure 1(b) shows the timing diagram representation of hold time.

## IV. FPGA VS. ASIC TIMING CHARACTERISTICS

FPGAs feature a structured architecture composed of Configurable Logic Blocks (CLBs), interconnect matrices, and dedicated routing resources. Their timing depends on logic block structure (LUTs with fixed delays, flip-flops with defined setup/hold times, carry chains, block RAM, DSP slices), routing architecture (programmable interconnect switches, hierarchical routing, parasitic effects), and clock distribution

TABLE I: Targeted platforms for experiments and their features

| | Target Platforms | |
|---|---|---|
| | Xilinx Kintex UltraScale+ (XCKU040) | ASIC 7nm FinFET Process |
| Specifications | Technology Node: 20nm planar process<br>Logic Cells : 537,600<br>Block RAM : 75.9 Mb<br>DSP Slices : 1,920<br>Clock Management Tiles : 6<br>Maximum CLK : 891 MHz (Typ) | Technology Node: 7nm FinFET<br>Standard Cell Library:<br>High-performance (HVT/SVT/LVT)<br>Metal Layers: 15<br>Minimum Feature Size: 7nm<br>Supply Voltage: 0.75V (nominal) |
| Key Timing Parameters | CLB Flip-Flop Setup : 180 ps<br>CLB Flip-Flop Hold : 120 ps<br>Clock-to-Q Delay : 450 ps<br>LUT6 Propagation Delay : 320 ps<br>Routing segment Delay : 200-800 ps | DFF Setup Time (SVT) : 45 ps<br>DFF Hold Time (SVT) : 35 ps<br>Clock-to-Q Delay (SVT) : 85 ps<br>NAND2 Propagation : 25 ps<br>Wire Delay (M1) : 15 ps/μm |

(balanced clock trees, Clock Management Tiles, regional/global networks) [26], [27].

ASICs use custom-designed logic gates and routing, relying on standard cell libraries with optimized gates, multiple drive strengths, and PVT-characterized cells. Custom routing involves carefully optimized metal layers, minimal parasitic, and custom clock tree synthesis.

Timing verification for both platforms is primarily based on Static Timing Analysis (STA), with FPGA- and ASIC-specific adaptations [28], [29]. Key STA components include [30], [31]:

**Path Analysis**: Setup paths consider launch flop clock-to-Q delay, combinational and routing delays, and capture flop setup time. Hold paths consider minimum clock-to-Q delay, minimum combinational and routing delays, and capture flop hold time. **Timing Constraints**: Clock definitions, input/output delays, false and multicycle path exceptions, and clock domain crossing constraints. FPGA timing analysis uses vendor-specific tools such as **Xilinx Vitis/Vivado** (place- and-route timing estimation, post-implementation verification, timing optimization) and **Intel Quartus TimeQuest** (slack analysis, critical path identification, optimization guidance). ASIC timing employs industry-standard tools like **Synopsys PrimeTime** (comprehensive STA, advanced delay models, crosstalk/noise analysis) and **Cadence Tempus** (parallel timing analysis, physical effects modeling, power-aware timing).

## V. CASE STUDY: XILINX KINTEX ULTRASCALE+ FPGA VS 7NM ASIC

Timing closure experiments were conducted on two platforms: the Xilinx Kintex UltraScale+ (XCKU040) FPGA and a 7nm FinFET ASIC. Table I summarizes the key features of these platforms.

### A. Test Circuit Implementation

A reference design—including a 32-bit ALU, 1024-entry register file, control state machine, clock domain crossing interfaces, and high-speed I/O—was implemented on both target platforms. Table II summarizes the resource utilization, physical implementation, and timing metrics for the FPGA and ASIC designs.

### B. Performance Comparison

The performance of both target platforms for an identical digital circuit implementation and measurement was evaluated as shown in Table III.

## VI. TIMING CLOSURE TECHNIQUES

Timing closure is a crucial challenge in modern digital design, requiring that timing constraints be met alongside functional correctness, power efficiency, and reliability. As performance demands grow and technology scales down, timing closure has shifted from a final step to an ongoing process throughout design, addressing variations in process, voltage, temperature, and complex clock domains [32], [33]. This section summarizes key timing closure techniques in FPGA and ASIC design, covering synthesis optimizations, place-and- route strategies, and constraint management. Effective timing closure depends on understanding hardware architecture, EDA tool capabilities, and carefully selecting and sequencing methods for maximum impact [11], [16], [23].

### A. FPGA Timing Closure Techniques

Timing closure in FPGA design relies on three core techniques: synthesis optimization, place and route optimization, and design constraints. The following sections explain each of these in detail.

**Synthesis Optimization** converts high-level descriptions into optimized gate-level netlists, setting the foundation for timing closure. It balances timing, area, and power through:

- *Register balancing and retiming*: Redistributes registers to equalize path delays and break critical paths, improving clock frequency [34].
- *Logic optimization and minimization*: Simplifies combinational logic to reduce delay, area, and power.
- *Resource sharing and pipelining*: Reuses hardware and segments long paths with registers to increase throughput.
- *Clock gating implementation*: Reduces dynamic power and clock skew by selectively disabling inactive logic [35].

**Place and Route Optimization** physically implements the design, directly affecting timing and power:

- *Floorplanning for Critical Paths* groups timing-critical blocks to minimize interconnect delays [36].
- *Placement Density Control* balances cell distribution to avoid routing congestion and excessive delays.

TABLE II: FPGA vs. ASIC Implementation Results

| Resource Type | Xilinx Kintex UltraScale+ (XCKU040) | | | ASIC 7nm FinFET Process | |
|---|---|---|---|---|---|
| | Utilized | Total | Utilization (%) | Parameter | Value |
| CLB LUTs | 45,230 | 242,400 | 18.7 | Core Area | 2.4 mm² |
| CLB Registers | 38,560 | 484,800 | 8.0 | Gate Count | 485,000 |
| Block RAM Tiles | 45 | 600 | 7.5 | Net Count | 512,000 |
| DSP Slices | 128 | 1,920 | 6.7 | Total Wire Length (mm) | 245 |
| **Timing Parameter** | **FPGA Value** | | | **ASIC Value** | |
| Maximum Clock Frequency (MHz) | 425 | | | 1250 | |
| Setup Slack (ns) | 0.125 | | | 0.045 | |
| Hold Slack (ns) | 0.089 | | | 0.028 | |
| Critical Path Delay (ns) | 2.35 | | | 0.80 | |
| **Critical Path Component** | **FPGA Delay (ns)** | | | **ASIC Delay (ns)** | |
| Register Clock-to-Q Delay | | 0.45 | | 0.085 | |
| Logic Delay | (4 LUT levels / 12 gate levels) | 1.28 | | 0.425 | |
| Routing / Wire Delay | | 0.62 | | 0.245 | |
| Setup Time Requirement | | 0.18 | | 0.045 | |

TABLE III: Performance and Design Flexibility Comparison Between FPGA and ASIC

| Timing Performance Metrics | | | | Design Flexibility Analysis | | |
|---|---|---|---|---|---|---|
| Metric | FPGA | ASIC | ASIC Advantage | Aspect | FPGA | ASIC |
| Max Frequency | 425 MHz | 1.25 GHz | 2.94× | Reconfigurability | Full runtime reconfiguration | Fixed post-fabrication |
| Setup Time | 180 ps | 45 ps | 4.0× | Time to Market | Days to weeks | Months to years |
| Hold Time | 120 ps | 35 ps | 3.43× | NRE Cost | $10K – $100K | $1M – $10M+ |
| Clock-to-Q Delay | 450 ps | 85 ps | 5.29× | Unit Cost (high volume) | $100 – $1,000 | $1 – $50 |
| Logic Delay per Level | 320 ps | 35 ps | 9.14× | Power Efficiency | 10–50× higher | Optimized |
| | | | | Performance | Good (100 MHz – 1 GHz) | Excellent (>1 GHz) |

- *Routing Congestion Management* detects and mitigates routing hotspots to preserve timing.
- *Clock Skew Optimization* ensures uniform clock arrival times, maximizing data path timing budgets.

**Design Constraints** provide precise timing, interface, and behavior specifications guiding synthesis and implementation tools. Accurate constraints enable timing-driven optimization and prevent timing violations or overly conservative designs.

```
# Clock
create_clock -period 2.35 [get_ports clk]
# I/O delays
set_input_delay -clock clk 0.5 \
  [get_ports data_in]
set_output_delay -clock clk 0.8 \
  [get_ports data_out]
# False path
set_false_path -from [get_clks clk_a] \
  -to [get_clks clk_b]
# Multicycle
set_multicycle_path -setup 2 \
  -from reg_a -to reg_b
```

The **Clock** constraint defines the primary clock on port *clk* with a 2.35 ns period (426MHz), establishing the timing reference for synchronous elements. The **I/O Delays** constraints set a 0.5 ns delay for the $data_{in}$ port relative to the clock edge, modeling external source delays and defining setup time for input registers. Also, the $data_{out}$ port must be stable 0.8 ns before the next clock edge, ensuring output timing meets downstream device requirements. The **False Path** constraint directs the timing analyzer to ignore paths crossing between asynchronous clock domains $clk_a$ and $clk_b$, reflecting non-critical or properly synchronized signals. The **Multicycle** path constraint allows the path between registers $reg_a$ and $reg_b$ to complete in 2 clock cycles, relaxing timing for multi-cycle operations or pipeline stages.

B. *Timing Closure Techniques for ASIC*

Timing closure in ASIC design involves synthesis optimization, physical design optimization, and advanced techniques, each addressing timing, power, and area constraints throughout the design flow [37], [38].

**Synthesis Optimization** differs from FPGA synthesis in its flexibility and complexity, leveraging techniques like:

- *Cell Sizing and Threshold Voltage Selection*: Standard cell libraries provide multiple drive strengths and $V_{th}$ options (HVT, RVT, LVT). Optimizers assign these based on timing criticality, balancing speed and leakage using statistical timing analysis.
- *Logic Restructuring and Boolean Optimization*: Advanced logic transformations—such as factorization, de-composition, and compound gate usage—reduce logic depth and delay. Tools also use ML for restructuring and technology mapping to minimize critical path delays. Key techniques include:
  - **Boolean manipulation**: Factorization, decomposition, and substitution help reduce logic depth and improve timing.
  - **Complex gate usage**: ASIC libraries often include multi-input and compound gates unavailable in FPGAs, allowing the optimizer to combine simple gates into faster, more efficient complex gates.
  - **Logic structure trade-offs**: The optimizer balances timing, area, and power by, for example, splitting large fan-in gates into cascaded smaller gates or merging multiple gates into compound functions that lower critical path delays.
  - **Physical considerations**: Restructuring also targets routing congestion and power distribution impacts, optimizing for better physical implementation.

- **Technology mapping**: Algorithms select optimal gate combinations from the standard cell library based on timing and area targets.
- **Don't-care optimization**: Unused input combinations are exploited to simplify logic further.
- **Machine learning assistance**: Modern tools use ML to discover and apply effective restructuring patterns across large designs.

- *Clock Tree Synthesis Optimization*: ASICs require custom clock networks (e.g., H-trees) with controlled skew, buffer placement, and integration of clock gating. Statistical analysis ensures robustness under process variations. Key points include:
  - **Custom clock tree design**: ASICs require tailored topologies (e.g., H-trees, meshes) based on layout and timing, unlike fixed FPGA networks.
  - **Buffer sizing and placement**: Buffers are optimized for strength and efficiency to ensure reliable clock delivery with minimal power/area.
  - **Skew control**: Skew is minimized to preserve timing margins across registers.
  - **Statistical robustness**: Clock trees are validated with statistical analysis to ensure reliability under process and environmental variations.
  - **Useful skew**: Intentional skew insertion can ease timing on critical paths by borrowing slack from non-critical ones.
  - **Clock gating**: Integrated gating logic disables clocks in idle regions for power savings without impacting timing.

  Overall, ASIC clock tree synthesis optimization balances timing precision, power efficiency, and manufacturability to support high-performance and low-power designs.

- *Power Optimization Techniques*: ASIC synthesis reduces dynamic power by minimizing switching activity via logic restructuring, operand isolation, and automatic clock gating (coarse to hierarchical). Static power, mainly from leakage, is managed using high-$V_{th}$ cells in non-critical paths and low-$V_{th}$ cells where timing is tight. Power gating cuts leakage by disconnecting idle blocks, requiring careful state retention. Multi-voltage designs assign supply levels based on performance needs. These strategies collectively reduce power while preserving timing across conditions.

**Physical Design Optimization** is the final and most detailed ASIC implementation stage, transforming the synthesized netlist into manufacturable silicon while ensuring timing closure. With full parasitic and process data now available, precise adjustments account for real-world effects—especially critical at advanced nodes where interconnect delays dominate and variations (e.g., lithography, metal density, stress) impact timing and reliability. Power integrity, thermal balance, and signal integrity must also be maintained. Modern tools use co-optimization across timing, power, area, and manufacturability, iteratively refining placement, routing, and timing. Techniques like gate sizing, buffer insertion, layer selection, and congestion relief are applied with detailed physical awareness, enabling fine-grained control over delays and resource usage. Coordinated optimization is essential for delivering correct, high-performance silicon.

- *Floorplanning and Placement Optimization* in ASIC design determines the optimal arrangement of standard cells and macros to minimize interconnect delay and routing congestion while satisfying timing, power, and manufacturing constraints. It begins with macro placement—positioning large blocks like memories and ana- log IP—to establish the chip's architectural structure. Standard cell placement follows, employing algorithms that reduce wirelength and prioritize timing-critical paths. Modern tools use analytical models for continuous optimization before legalizing cell positions to manufacturing grids. Timing-driven placement addresses setup and hold requirements, reduces clock skew, and optimizes data path delays. Early parasitic estimation incorporates routing and via effects, enabling detection of potential timing issues. Placement also ensures power integrity by maintaining adequate IR drop margins and supply connectivity. Congestion-aware techniques balance placement density to prevent routing bottlenecks and improve routability. Overall, placement integrates timing, power, and physical constraints to support efficient, reliable implementation.

- *Clock Tree Synthesis and Balancing* in ASIC physical design constructs the physical clock distribution network to meet stringent skew, power, and timing targets. Unlike logical synthesis, this stage implements buffers and interconnects accounting for parasitic, manufacturing variation, supply noise, and temperature gradients. It starts with clustering clock sinks based on proximity and timing needs, then builds hierarchical buffer trees for each cluster. Buffer sizing and placement optimize drive strength, power efficiency, and area usage. Routing is distributed across layers and vias to balance performance and re- source use. Useful skew techniques—controlled clock arrival variations—help relax setup or hold constraints on critical paths, coordinated closely with timing analysis. Clock gating structures are integrated to enable dynamic power management without compromising timing. Post- synthesis refinements such as buffer resizing, rerouting, and tree restructuring are driven by detailed parasitic extraction and timing feedback to ensure timing closure.

- *Routing Optimization for Timing* focuses on minimizing signal delays during interconnect routing while adhering to design rules and maximizing routing resource efficiency. With full placement data, algorithms optimize routing topology, layer usage, and via placement to improve timing on critical nets. The flow starts with global routing to define coarse paths and layer assignments, followed by detailed routing to finalize wire geometries. Timing-driven routing prioritizes critical nets, applying techniques like layer promotion, buffer insertion, and path shaping to reduce delays. Electrical effects such as crosstalk, RC delay, and inductance are accounted for, especially at advanced nodes. The process includes concurrent buffer sizing and placement to reduce delays on long wires and select routing topologies that balance performance and manufacturability. Post-route refinements like wire widening, shielding, and topology adjustments use precise timing analysis. Design rules for spacing, via use, and metal fill are strictly enforced to ensure performance and yield.

- *Post-Route Timing Optimization* is the final step toward timing closure, leveraging full parasitic and layout data to meet performance goals. Detailed timing analysis identifies residual violations and improvement of opportunities. Localized changes—gate resizing, buffer insertion or resizing, and minor routing adjustments—fix setup and hold issues without compromising overall design integrity. Useful skew optimization further relaxes timing constraints, while netlist edits like pin swapping or gate replacement address bottlenecks. Power optimization exploits timing slack to reduce consumption without impacting timing. Throughout, manufacturing design rule checks ensure yield and compliance. The process concludes with exhaustive timing verification across all process corners and modes, guaranteeing robust closure before tape-out.

**Advanced Timing Closure Techniques in ASICs** address the most challenging timing issues that arise when traditional optimization methods become insufficient, especially in high-performance or deeply scaled technologies. These techniques exploit fine-grained circuit behaviors, process nuances, and design flexibilities not covered by standard flows, requiring deep expertise in circuit design, physics, and algorithmic optimization. As designs move into nanometer regimes, secondary effects such as process variation, voltage droop, and temperature dependence increasingly dominate timing margins. Advanced timing methods must accurately model these factors and perform careful trade-off analyses, as even minor adjustments can significantly impact performance or reliability. These strategies solve complex multi-variable optimization problems balancing timing, power, area, and manufacturability simultaneously. The tools employed combine precise models with scalable algorithms—often relying on heuristics or ap- proximations to meet runtime constraints. Success depends on tight integration with characterization data and robust signoff verification. The following specialized, state-of-the-art techniques target the toughest timing bottlenecks; each method addresses specific scenarios and is often combined with others for maximum effect. These approaches are essential for ASICs at the edge of performance and technology scaling.

- *Useful Skew Insertion* intentionally introduces controlled clock arrival differences (skews) between sequential elements to improve timing closure. Unlike conventional zero-skew clock trees, this technique strategically shifts clock edges to relax setup constraints on critical paths by borrowing slack from less critical ones. It begins with detailed timing analysis to identify paths that benefit from relaxed timing and those that tolerate tighter constraints. An optimization algorithm computes ideal skew values, balancing setup and holds requirements while considering clock tree implementation limits. Effective deployment requires precise clock tree synthesis capable of delivering targeted skews with minimal variation and must account for manufacturing variability and environmental effects. In advanced designs, adaptive skew networks dynamically adjust skew during runtime to maintain timing across changing conditions. This technique is especially valuable in high-speed processors, where slight timing gains on critical paths translate into significant performance improvements.

- *Gate Sizing Optimization* adjusts transistor dimensions of individual logic gates to finely control timing and power. Unlike fixed standard cell selections, this method enables continuous tuning of drive strength within fabrication limits, achieving performance and efficiency gains beyond standard libraries. Advanced algorithms—often gradient-based or evolutionary—model the impact of gate sizing on delay and power, optimizing transistor widths while considering switching speed, power consumption, and manufacturability. Guided by detailed timing analysis, adjustments improve critical paths without unnecessary area or power increases. Modern approaches integrate statistical techniques to ensure robustness across process corners and temperature variations, accounting for worst- case scenarios to maintain reliability post-fabrication. The method also considers layout density and power delivery, as upsized gates increase capacitance and current demand. Gate sizing is particularly effective when combined with useful skew insertion and custom layouts, forming a synergistic strategy for high-performance timing closure.

- *Buffer Insertion Strategies* optimize signal propagation by placing buffers along interconnects to reduce delay caused by wire resistance and capacitance—critical at advanced nodes where interconnect delay dominates gate delay. The goal is to improve timing without incurring excessive power or area overhead. Dynamic programming or similar algorithms evaluate possible buffer placements and sizes, balancing delay reduction with buffer cost to meet timing targets efficiently. Physical design constraints limit buffer locations and types, adding complexity. Mod- ern multi-objective optimization tailors buffer choices to local needs, prioritizing speed on critical paths and energy efficiency elsewhere. Statistical analysis ensures timing robustness under process variations. Buffer insertion is tightly integrated with routing optimization, allowing simultaneous tuning of wire topology and buffer placement to preserve timing improvements through final layout, significantly aiding timing closure in dense, high-performance ASICs.

- *Wire Sizing and Spacing Optimization* fine-tunes interconnect electrical properties—resistance, capacitance, and crosstalk—by varying wire widths and spacing within manufacturing constraints and routing density limits. Unlike uniform routing, this technique customizes wire dimensions to minimize delay and power without com- promising signal integrity or manufacturability. Detailed electrical analyses identify optimal widths: wider wires reduce resistance but increase capacitance and routing congestion, so the optimization balances these trade-offs. Crosstalk is addressed via spacing adjustments and joint optimization across neighboring nets using electromagnetic coupling models. All designs respect manufacturing rules for minimum widths and spacing to ensure yield and reliability. Advanced methods include statistical analysis to mitigate electromigration and defect risks in narrower wires, balancing performance gains with manufacturing robustness. This technique is crucial for high-performance designs and integrates with sophisticated layout tools and verification flows to meet timing, power, and manufacturability goals.

## VII. Advanced Timing Considerations in FPGA and ASIC Design

Modern digital circuit design increasingly faces complex timing challenges driven by technology scaling and growing performance demands. Traditional deterministic timing models, sufficient for older nodes, are inadequate at nanometer scales where process variations, voltage fluctuations, temperature gradients, and aging induce statistical timing behaviors. Addressing these issues requires advanced analysis methods and robust design techniques.

This section explores timing considerations beyond conventional static timing analysis, emphasizing the stochastic nature of modern circuits and specialized approaches for reliable operation. It covers both FPGA and ASIC platforms, highlighting differences in how process variations affect each and exploring corresponding mitigation strategies. Key challenges such as clock domain crossing in multi-clock system-on-chip designs are also examined. Combining current research with practical design guidelines, the content offers quantitative analyses, methodology recommendations, and validation techniques geared toward industrial applications.

### A. Process Variation in FPGAs

FPGAs face unique challenges due to their uniform architecture and statistical usage of many replicated resources. Unlike ASICs—where variations uniformly affect specific custom circuit elements—FPGA variations manifest as statistical distributions across numerous identical components. This section analyzes key variation mechanisms impacting FPGA timing and provides quantitative guidelines for robust design under these conditions.

*a) Manufacturing Variations in LUT Delays:* Variations arise from threshold voltage shifts, channel length changes, and oxide thickness differences in transistors. Measurements on 28nm FPGAs show LUT delay standard deviations of 8–12

**Design recommendations:**
1) Target synthesis constraints at 85
2) Use FPGA-specific statistical timing analysis tools when available.
3) Reduce logic depth for timing-critical paths exceeding 20 LUT levels.
4) Analyze timing slack distribution to identify paths most vulnerable to variation-induced failures.

*b) Interconnect Resistance and Capacitance Variations:* Variations in metal thickness, dielectric constants, and via resistance cause delay standard deviations of 6–10

**Design recommendations:**
1) Maintain logic utilization below 70
2) Apply hierarchical design methods to reduce global interconnect needs.
3) Use relative placement constraints to shorten critical path routing distances.
4) Insert pipelines for paths spanning more than 4–6 routing hierarchy levels.

*c) Temperature and Voltage Effects:* FPGA timing is sensitive to temperature (delay changes of -1.5

**Design recommendations:**
1) Apply 15–20
2) Use temperature-aware placement for timing-critical logic.
3) Employ dynamic voltage scaling to mitigate temperature-induced timing shifts.
4) Integrate environmental monitoring and adaptive timing techniques when feasible.

*d) Aging-Induced Degradation:* FPGA timing degrades over time due to bias temperature instability (BTI), hot carrier injection (HCI), and time-dependent dielectric breakdown (TDDB). Timing degradation of 2–5

**Design recommendations:**
1) Include aging-aware timing margins of 3–7
2) Incorporate monitoring circuits to detect aging-induced timing shifts.
3) Consider design refresh strategies for critical applications.
4) Minimize stress-inducing operating conditions to slow aging.

### B. Process Variation in ASICs

ASIC process variations differ due to custom layouts and diverse circuit elements. Variation impact depends heavily on transistor sizes, layout patterns, and local density. This section provides quantitative insights and design strategies for robust ASIC timing.

*a) Within-Die (WID) Variations:* Systematic gradients and random fluctuations cause spatially correlated variations. At 7nm nodes, threshold voltage varies 4–8

**Design recommendations:**
1) Cluster critical paths to maximize correlation benefits.
2) Use variation-aware placement that models spatial correlation.
3) Apply statistical timing analysis incorporating correlation.
4) Employ adaptive body biasing to compensate systematic variation.

*b) Die-to-Die (D2D) Variations:* Global parameter shifts affect entire dies uniformly, with 8–15

**Design recommendations:**
1) Use adaptive timing techniques to compensate for D2D variation.
2) Employ post-silicon tuning, e.g., adaptive voltage scaling.

3) Perform multi-corner timing closure accounting for D2D-WID correlations.
4) Apply statistical binning to optimize yield across variation ranges.

*c) Process Corner Analysis:* Traditional corners (SS, TT, FF) estimate timing bounds efficiently but struggle with nanometer node complexities and non-Gaussian, multi-modal distributions. Corner methods may underestimate timing failures by 2–5× compared to statistical timing. They remain useful for early closure and conservative validation if carefully calibrated.

**Design recommendations:**
1) Use corner analysis for initial closure with appropriate guard bands.
2) Apply statistical timing verification for final sign-off.
3) Incorporate intermediate corners (SF, FS) for better coverage.
4) Calibrate corner results against statistical timing data.

## VIII. CLOCK DOMAIN CROSSING AND SYNCHRONIZATION TECHNIQUES

Clock Domain Crossing (CDC) is a critical challenge in modern SoC design, where multiple clock domains with differ- ent frequencies and phases coexist. The increasing complexity of multi-core processors and mixed-signal integration has significantly heightened CDC issues. Improper handling of CDC can cause metastability, data corruption, and unpredictable system behavior, making robust synchronization and thorough verification essential. At its core, CDC failure arises from timing violations of setup and hold requirements when signals traverse asynchronous clock domains. Unlike single- clock systems, CDC circuits must handle uncertainty in signal arrival times relative to the destination clock edges, requiring probabilistic design methods and careful mean time between failures (MTBF) analysis to ensure reliability [29].

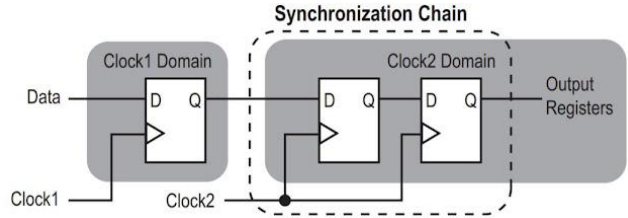

Figure 2. Synchronizing flip-flops configuration

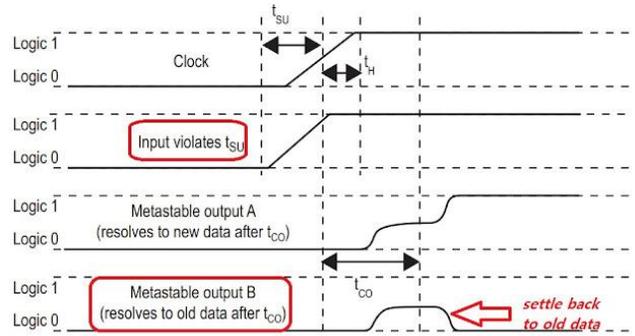

Figure 3. Preventing metastability using synchronizing flip-flops

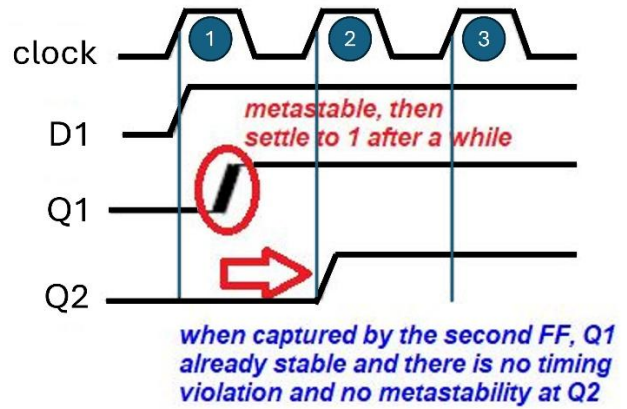

Figure 4. Preventing metastability in a simplified scheme

The two-flip-flop synchronizer is the fundamental technique for synchronizing single-bit signals crossing asynchronous clock domains. It uses two cascaded flip-flops clocked by the destination domain clock. The first flip-flop resolves metastability, while the second provides additional settling time, effectively trading a minimum two-cycle latency for greatly improved reliability. Figure 2 illustrates this configuration, while Figure *3* shows the timing diagram demonstrating how metastability is prevented using synchronizing flip-flops technique. Figure 4 further clarifies the timing behavior: if the first flip-flop (FF1) violates setup or hold times, its output (Q1) may enter a metastable state. After an uncertain clock-to-output delay ($T_{co}$), the output stabilizes. The second flip-flop (FF2) then samples this stabilized signal on the next clock edge, preventing metastability propagation downstream. Longer clock periods in the destination domain increase the available resolution time, further reducing metastability risk.

Due to its simplicity, effectiveness, and well-understood behavior, the two-flip-flop synchronizer remains the most widely used CDC synchronization method. However, advanced process nodes with lower supply voltages and increased transistor variability make metastability resolution more challenging. To mitigate this, custom flip-flops with enhanced feedback strength and optimized sizing are used to improve the metastability resolution time constant.

**Mean Time Between Failures (MTBF)** analysis provides

Modern FPGA and ASIC CDC designs face distinct challenges and advantages. FPGAs offer flexible clock allocation and built-in CDC primitives, but are constrained by
routing-induced skew and limited dedicated CDC resources. ASICs provide superior performance and power efficiency yet demand stricter validation and afford less post-fabrication flexibility. Both platforms benefit from advanced EDA tools that automate CDC verification and optimize cross-domain signal handling [39].

Synchronization techniques are key to reliable CDC, with each method presenting trade-offs in performance, area, and power. Modern systems often combine multiple synchronization methods, requiring deep understanding of their interactions. The rise of higher clock speeds, lower voltages, and stringent reliability requirements, especially in safety-critical designs—has driven the evolution of synchronization methods. Contemporary techniques also address electromagnetic interference (EMI), power noise, and process variation effects, incorporating adaptive mechanisms, self-test features, and real- time monitoring to maintain robustness under dynamic operating conditions.

**Two-Flip-Flop Synchronizers**

a quantitative reliability metric by linking metastability probability to available resolution time and flip-flop characteristics. The probability that metastability persists beyond the resolution time decays exponentially, so even small increases in resolution time yield large improvements in MTBF.

The fundamental MTBF relationship for a two-flip-flop synchronizer is:

$$\text{MTBF} = \frac{e^{\frac{T_{\text{res}}}{\tau}}}{f_{\text{data}} \times f_{\text{clock}} \times T_w} \quad (5)$$

where:

- $T_{\text{res}}$ = available resolution time (typically one clock period minus setup time).
- $\tau$ = metastability resolution time constant (technology-dependent).
- $f_{\text{data}}$ = asynchronous data transition frequency.
- $f_{\text{clock}}$ = destination clock frequency.
- $T_w$ = metastability window width (time during which a data transition can cause metastability).

Precise MTBF estimates require accurate characterization of $\tau$ and $T_w$, which vary with process, voltage, temperature, and device variations. Advanced MTBF models incorporate these statistical variations to yield confidence intervals rather than simple point estimates [39]–[41].

## IX. Practical Discussions in FPGA Design

This section presents hands-on discussions and experimental case studies based on topics covered in this paper, aimed at deepening understanding of timing closure in digital design and its importance. The code examples were implemented and tested on a Xilinx Kintex UltraScale+ FPGA (XCKU040) development board.

### A. Process Technology Impact on MTBF

Advanced process nodes introduce challenges for metastability analysis due to increased variations, lower supply volt- ages, and shorter intrinsic time constants. As process scales down, the time constant $\tau$ decreases, requiring longer resolution times to maintain equivalent MTBF. Supply voltage fluctuations significantly affect both $\tau$ and the metastability window $T_w$, making worst-case analysis across the full voltage range essential. Below is example Verilog code implementing an enhanced synchronizer with MTBF monitoring capability.

```verilog
// Enhanced synchronizer with MTBF monitoring
// capability
module mtbf_aware_sync #(
    parameter DEPTH = 2,
    parameter MTBF_TARGET = 1e12 // Target
    MTBF in hours
) (
    input   clk,
    input   rst_n,
    input   async_in,
    output  sync_out,
    output  mtbf_violation // Flag for MTBF
    monitoring
);

reg [DEPTH-1:0] sync_chain;
reg prev_sync_out;
wire metastability_detected;

always @(posedge clk or negedge rst_n)
begin
    if (!rst_n)
        sync_chain <= 0;
    else
        sync_chain <= {sync_chain[DEPTH-2:0],
        async_in};
end

assign sync_out = sync_chain[DEPTH-1];

// Metastability detection logic (simplified)
always @(posedge clk) begin
    prev_sync_out <= sync_out;
end

assign metastability_detected =
(sync_chain[DEPTH-1] !== sync_chain[DEPTH-2])
    && (prev_sync_out == sync_chain[DEPTH-1]);
assign mtbf_violation = metastability_detected;

endmodule
```

### B. Two-Flip-Flop Synchronizer

: FPGA-based two-flip-flop synchronizers leverage dedicated synchronizer primitives found in modern architectures like Xilinx UltraScale+ and Intel Stratix. These devices offer specialized flip-flops with enhanced metastability resolution and built-in timing analysis support. Proper placement and routing of synchronizer flip-flops is crucial to minimizing clock skew and maintain sufficient separation between stages. Designs should use dedicated clock networks and avoid shared routing resources to reduce crosstalk and timing variability. Below is example Verilog code implementing a two-flip-flop synchronizer:

```verilog
// FPGA-optimized two-flop synchronizer
(* ASYNC_REG = "TRUE" *)
(* SHREG_EXTRACT = "NO" *)
reg sync_ff1, sync_ff2;

always @(posedge dest_clk or negedge rst_n)
  begin
    if (!rst_n) begin
        sync_ff1 <= 1'b0;
        sync_ff2 <= 1'b0;
    end else begin
        sync_ff1 <= async_signal;
        sync_ff2 <= sync_ff1;
    end
end
```

### C. FIFO-Based Synchronization & Gray Code Pointer Management

: FIFO-based synchronization is an effective method for multi-bit data transfer across clock domains, preserving data integrity and order unlike single-bit two-flip-flop synchronizers. It uses separate read and writes pointers operating in their respective clock domains, with empty and full flags generated through Gray code pointer comparisons to avoid metastability during multi-bit synchronization. Asynchronous FIFOs typically implement dual-port memories to allow con- current read/write operations without conflicts, and the FIFO depth must handle maximum data rate differences and bursts. A key challenge is safely comparing pointers across clock domains. Binary counters risk metastability due to multiple simultaneous bit changes, whereas Gray code counters change only one bit per increment, mitigating this risk. Despite this, Gray code pointers still require two-flop synchronizers before cross-domain comparison to ensure stability. Below is example Verilog code implementing a Gray code counter for FIFO pointer management:

```verilog
// Gray code counter for FIFO pointer
management
always @(posedge clk or negedge rst_n)
begin
    if (!rst_n)
        gray_ptr <= 0;
    else if (enable)
        gray_ptr <= (gray_ptr >> 1) ^
        gray_ptr
        + 1;
end

// Pointer synchronization and comparison
always @(posedge rd_clk) begin
    wr_ptr_sync1 <= wr_gray_ptr;
    wr_ptr_sync2 <= wr_ptr_sync1;
    empty <= (rd_gray_ptr == wr_ptr_sync2);
end
```

*D. Handshaking Protocols Implementation*

: Handshaking protocols enable reliable data transfer between asynchronous clock domains through explicit request- acknowledge exchanges, making them ideal for single-word transfers and applications requiring precise confirmation. Unlike FIFO-based methods that use buffering, handshaking coordinates transfer timing with control signals to ensure data integrity. Four-phase handshaking, the most robust variant, in- volves a complete cycle of request assertion, acknowledgment, request de-assertion, and acknowledgment de-assertion, offering deterministic operation across varying clock frequencies and preventing race conditions. Two-phase protocols provide higher throughput but require more complex design to avoid timing hazards. The choice depends on system requirements, performance needs, and clock domain characteristics. Implementing four-phase handshaking requires careful synchronization of control signals to avoid metastability and race conditions. The typical sequence is: request asserted with data, receiver asserts acknowledgment after capturing data, sender de-asserts request, and receiver de-asserts acknowledgment to complete the cycle. Below is example Verilog code illustrating a four-phase handshaking transmitter and receiver implementation [42].

```verilog
//Four-phase handshaking transmitter
typedef enum {IDLE, WAIT_ACK, WAIT_ACK_LOW} tx_state_t;
tx_state_t tx_state;

always @(posedge tx_clk or negedge rst_n)
begin
    if (!rst_n)
    begin
        tx_state <= IDLE;
        tx_req <= 1'b0;
        tx_data_reg <= 0;
    end else begin
        case (tx_state)
            IDLE: if (data_valid)
            begin
                tx_data_reg <= data_in;
                tx_req <= 1'b1;
                tx_state <= WAIT_ACK;
            end
            WAIT_ACK: if (ack_sync)
            begin
                tx_req <= 1'b0;
                tx_state <= WAIT_ACK_LOW;
            end
            WAIT_ACK_LOW: if (!ack_sync)
            begin
                tx_state <= IDLE;
            end
        endcase
    end
end

// Four-phase handshaking receiver
typedef enum {IDLE, DATA_VALID, WAIT_REQ_LOW} rx_state_t;
rx_state_t rx_state;

always @(posedge rx_clk or negedge rst_n)
begin
    if (!rst_n) begin
        rx_state <= IDLE;
        rx_ack <= 1'b0;
        data_out_reg <= 0;
        data_ready <= 1'b0;
    end else begin
        case (rx_state)
            IDLE: if (req_sync)
            begin
                data_out_reg <= tx_data_reg;
                data_ready <= 1'b1;
                rx_ack <= 1'b1;
                rx_state <= DATA_VALID;
            end
            DATA_VALID: if (data_consumed)
            begin
                data_ready <= 1'b0;
                rx_state <= WAIT_REQ_LOW;
            end
            WAIT_REQ_LOW: if (!req_sync)
            begin
                rx_ack <= 1'b0;
                rx_state <= IDLE;
            end
        endcase
    end
end
```

*E. Gray Code Counters & Binary-to-Gray Conversion Implementation*

: Gray code counters provide essential infrastructure for safe multi-bit signal transfer across clock domain boundaries by ensuring only one bit changes per count increment, eliminating the possibility of transient invalid states during asynchronous sampling. The mathematical properties of Gray code sequences make them ideally suited for pointer management in asynchronous FIFOs, address generation for dual-port memories, and any application requiring monotonic multi-bit signal progression across CDC boundaries. The implementation of Gray code counters require careful consideration of the conversion between binary and Gray code representations and the implications for downstream logic that must operate on the encoded values. The design of efficient Gray code counters involve optimization of the conversion logic to minimize propagation delay and power consumption while maintaining the essential single-bit-change property. Advanced implementations may incorporate parallel Gray code generation for high-speed applications and specialized decoding logic for applications requiring both Gray and binary representations. The verification of Gray code counter implementations require specialized testbenches that verify the single-bit-change property across all possible state transitions and boundary conditions. The conversion from binary to Gray code follows the mathematical relationship where each Gray code bit is the XOR of the corresponding binary bit and the next higher-order binary bit. This conversion can be implemented efficiently using simple XOR gates, making it suitable for high-speed applications. The most significant bit of the Gray code equals the most significant bit of the binary code, simplifying the con- version logic. A piece of example Verilog code to implement an Efficient binary-to-Gray code converter is shown below.

```verilog
// Efficient binary-to-Gray code converter
```

```verilog
function automatic [WIDTH-1:0] bin_to_gray
(input [WIDTH-1:0] binary);
    bin_to_gray = binary ^ (binary >> 1);
endfunction

// Gray code counter with integrated
conversion
always @(posedge clk or negedge rst_n)
begin
    if (!rst_n) begin
        binary_count <= 0;
        gray_count <= 0;
    end else if (enable)
    begin
        binary_count <= binary_count + 1;
        gray_count <= bin_to_gray
        (binary_count+ 1);
    end
end
```

### F. Synchronizer Depth Requirements & Adaptive Depth Control Mechanisms

Determining the appropriate synchronizer depth is crucial to balancing reliability with latency and power constraints. While two-flop synchronizers suffice for most cases, safety-critical and high-speed applications often require additional stages, as MTBF improves exponentially with depth. Advanced analysis also accounts for transient conditions like power-up, frequency changes, and environmental stress. Adaptive synchronizers dynamically adjust their depth based on real-time metastability measurements and operating conditions, optimizing reliability without excessive performance penalties. Implementing such variable-depth synchronizers demands complex control logic and thorough verification to ensure correctness across all configurations. These systems typically use statistical methods, like exponential smoothing, to distinguish transient glitches from sustained metastability trends and adjust synchronizer depth accordingly.

```verilog
// Adaptive depth synchronizer with real-time
optimization
module adaptive_sync #(
    parameter MAX_DEPTH = 5,
    parameter MIN_DEPTH = 2)
(
    input   clk,
    input   rst_n,
    input   async_in,
    input   [2:0] reliability_mode,
    // 0=performance, 7=max_reliability
    output  sync_out,
    output  [2:0] current_depth);

reg [MAX_DEPTH-1:0] sync_chain;
reg [2:0] active_depth;
reg [15:0] metastability_counter;
reg [23:0] mtbf_timer;

// Depth selection based on reliability
mode and history
always @(posedge clk or negedge rst_n)
begin
    if (!rst_n) begin
        active_depth <= MIN_DEPTH;
        metastability_counter <= 0;
        mtbf_timer <= 0;
    end else begin
        // Update MTBF monitoring
        if (mtbf_timer == 24'hFFFFFF)
        begin
            // Evaluate metastability
            rate and adjust depth
            if (metastability_counter
            > reliability_mode)
            begin
                if (active_depth <
                MAX_DEPTH)
                    active_depth <=
                    active_depth + 1;
            end
            else if
            (metastability_counter == 0)
            begin
                if (active_depth >
                MIN_DEPTH)
                    active_depth <=
                    active_depth - 1;
            end
            metastability_counter <= 0;
            mtbf_timer <= 0;
        end else begin
            mtbf_timer <= mtbf_timer + 1;
        end
    end
end

// Synchronizer chain with variable
tap selection
always @(posedge clk or negedge rst_n)
begin
    if (!rst_n)
        sync_chain <= 0;
    else
        sync_chain <=
        {sync_chain[MAX_DEPTH-2:0], async_in};
end

assign sync_out = sync_chain[active_depth-1];
assign current_depth = active_depth;

endmodule
```

### G. Clock Frequency Relationships & Dynamic Frequency Scaling

The clock frequency relationship between communicating domains critically influences CDC design and performance. Rational frequency ratios (where one frequency is an integer multiple of the other) allow deterministic timing analysis and simpler synchronizer designs, while irrational ratios demand probabilistic methods and larger safety margins. These ratios affect data throughput, buffering needs, and flow control complexity. High frequency ratios complicate timing closure and may require techniques like frequency division or phase interpolation, whereas very low ratios increase buffering and latency. Modern CDC design leverages frequency planning to optimize domain relationships and reduce synchronization complexity. Dynamic frequency scaling further complicates CDC design by introducing time-varying frequency ratios. Adaptive synchronizers must handle these changes seamlessly, often require frequency change notifications and temporary synchronizer adjustments during transitions to maintain data integrity and avoid metastability. An example of Verilog implementation of a frequency-aware CDC controller with adaptive timing is provided below.

```verilog
// Frequency-aware CDC controller
with adaptive timing
module freq_aware_cdc #(
    parameter MAX_FREQ_RATIO = 16
) (
    input src_clk,
    input dst_clk,
    input rst_n,
    input [3:0] freq_ratio,
    // Current frequency ratio
    input freq_change_req,
    // Frequency change notification
    input async_data_in,
    output sync_data_out,
    output transfer_ready
);

reg [3:0] current_ratio;
reg freq_change_pending;
reg [2:0] sync_depth;
reg transfer_enable;

// Adaptive synchronizer depth based on
frequency ratio
always @(*) begin
    case (freq_ratio)
        4'd1: sync_depth = 3'd2;
        // 1:1 ratio - minimum depth
        4'd2, 4'd3, 4'd4: sync_depth =
        3'd3; // Low ratios
        default: sync_depth = 3'd4;
        // High ratios - maximum depth
    endcase
end

// Frequency change handling
always @(posedge dst_clk or negedge rst_n)
begin
    if (!rst_n) begin
        current_ratio <= 4'd1;
        freq_change_pending <= 1'b0;
        transfer_enable <= 1'b1;
    end else begin
        if (freq_change_req && !
        freq_change_pending) begin
            freq_change_pending <= 1'b1;
            transfer_enable <= 1'b0;
            // Disable during transition
        end else if (freq_change_pending)
        begin
            current_ratio <= freq_ratio;
            freq_change_pending<=1'b0;
            transfer_enable <= 1'b1;
        end
    end
end

assign transfer_ready = transfer_enable
&& !freq_change_pending;

endmodule
```

## X. TECHNOLOGY SCALING IMPACT ON FPGA & ASIC DESIGN

As technology scales to 7nm, 5nm, and beyond, digital design faces growing challenges that reshape timing closure, power management, and reliability strategies for both FPGAs and ASICs. Process variations intensify, with greater threshold voltage fluctuations relative to supply voltage, reducing timing margins and demanding more conservative constraints in FPGAs and advanced statistical timing analysis in ASICs. Setup and hold times tighten as clock frequencies approach physical limits, while interconnect delays increasingly dominate over gate delays, shifting design focus from gates to interconnect optimization. Power consumption, especially leakage current, becomes a critical bottleneck, with thermal runaway risks requiring careful circuit and system-level mitigation. Process variations arise from systematic manufacturing differences, random atomic-level effects, and aging mechanisms like BTI and HCI, all accumulating over time. Designs must therefore ensure robust functionality across initial manufacturing variability and throughout device lifetimes.

Figure 5 represents the technology scaling impact visualization diagram. The exponential curve for Process Variation Trends (top left graph) shows process variations increasing dramatically as technology nodes shrink from 180nm to 22nm as follows:

- 180nm node: 5% variation (relatively manageable)
- 90nm node: 10% variation (doubling of uncertainty)
- 45nm node: 15% variation (3x increase from 180nm)
- 22nm node: 20% variation (4x increase from baseline)

This exponential growth means that at advanced nodes, the same circuit design will have much wider performance distributions, making timing closure significantly more challenging. The stacked bar chart for power consumption evolution (top right) reveals a critical shift in power consumption patterns as follows:

- 180nm: Dynamic power (blue) dominates, with minimal static power (red)
- 90nm: Static power begins to emerge as a significant component
- 45nm: Static and dynamic power become roughly equal
- 22nm: Static power actually exceeds dynamic power

This crossover point represents a fundamental change in design priorities - leakage current becomes the primary power concern rather than switching activity.

Finally, The shrinking bars for the timing margin degradation (bottom section) dramatically illustrate how timing margins erode with scaling as follows:

**Setup Margins**: Show severe degradation from a comfortable 200-unit margin at 180nm down to just 60 units at 22nm - a 70% reduction. **Hold Margins**: Similarly degrade from 100 units to 40 units, representing a 60% reduction. The arrows connecting setup and hold margins emphasize that both critical timing parameters are simultaneously under pressure.

The key aspects of engineering implications derived from Figure 5 are summarized as follows:

1) Design Methodology Impact: Traditional corner-based analysis becomes inadequate; statistical timing analysis becomes essential at advanced nodes.

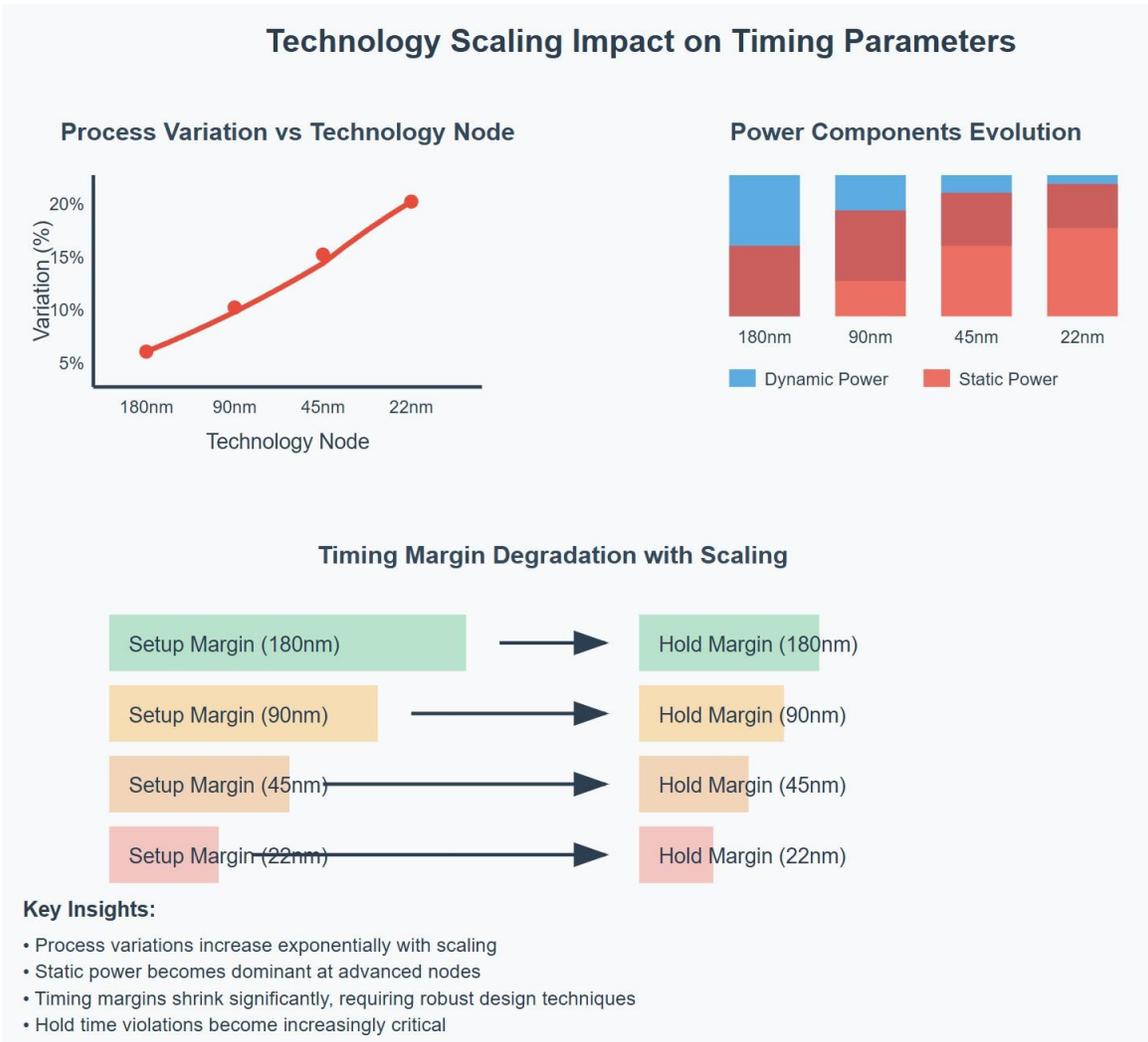

Figure 5. Technology scaling impact visualization diagram

2) Power Management: The dominance of static power requires new approaches like power gating, multi-threshold libraries, and dynamic voltage scaling.
3) Timing Closure: Shrinking margins means designs must be more conservative, requiring larger guard bands and more sophisticated optimization techniques.
4) Yield Considerations: Higher process variations directly impact manufacturing yield, necessitating yield-aware design methodologies.

This effectively demonstrates why advanced node design requires fundamentally different approaches compared to older technologies - the traditional design assumptions no longer hold when variations become the dominant factor affecting circuit behavior.

**Recommendations**: For FPGA designs, employ adaptive timing methods that adjust for process variations via runtime reconfiguration. Use built-in process monitors to track local conditions and dynamically refine timing constraints. Adopt margin-aware placement and routing strategies that balance worst-case corner considerations with typical operating conditions. ASIC designs should integrate statistical static timing analysis (SSTA) early, moving beyond fixed corner analysis. Leverage adaptive voltage scaling (AVS) and adaptive frequency scaling (AFS) to optimize power and performance amid process variability. Incorporate aging models into timing tools to ensure long-term reliability. Both FPGA and ASIC workflows benefit from machine learning–based predictive timing analysis, utilizing historical data to enhance closure efficiency. Cross-corner optimization techniques that handle multiple process corners simultaneously should be implemented to achieve robust, efficient designs.

## XI. CONCLUSION

This comprehensive analytical survey highlights the significant differences in setup and hold timing characteristics between FPGAs and ASICs, demonstrating ASICs' superior timing performance with setup and hold times of 45ps and 35ps, respectively, compared to 180ps and 120ps for modern FPGAs. Despite this, FPGAs remain competitive due to their design flexibility and faster time-to-market. ASICs deliver 3-4× better timing, greater power efficiency, and more predictable performance across process variations. FPGAs, however, offer lower non-recurring engineering costs but higher per-unit costs on a scale. Choosing between FPGA and ASIC depends on application needs, production volume, deadlines, and performance targets. Understanding these timing differences is essential for effective timing closure and design optimization. Future work should explore advanced timing optimizations, effects of emerging technologies, and automated design methodologies to help close the performance gap between FPGA and ASIC implementations.